\documentclass[%
preprint,
superscriptaddress,
 amsmath,amssymb,
 aps,
 prx,
]{revtex4-2}

\usepackage{chemformula} % Formula subscripts using \ch{}
\usepackage{xfrac}  % Works better with other fonts
\usepackage{float} 
\usepackage{physics}

\usepackage{graphicx}% Include figure files
\usepackage{dcolumn}% Align table columns on decimal point
\usepackage{bm}% bold math
\begin{document}

\title{Probing the coherence of solid-state qubits at avoided crossings}% Force line breaks with \\

\author{Mykyta Onizhuk}
\affiliation{Department of Chemistry, University of Chicago, Chicago, IL 60637, USA}
\affiliation{Pritzker School of Molecular Engineering, University of Chicago, Chicago, IL 60637, US}
\author{Kevin C. Miao}
\affiliation{Pritzker School of Molecular Engineering, University of Chicago, Chicago, IL 60637, US}
\author{Joseph P. Blanton}
\affiliation{Department of Physics, University of Chicago, Chicago, IL 60637, USA}
\affiliation{Pritzker School of Molecular Engineering, University of Chicago, Chicago, IL 60637, US}
\author{He Ma}
\affiliation{Department of Chemistry, University of Chicago, Chicago, IL 60637, USA}
\affiliation{Pritzker School of Molecular Engineering, University of Chicago, Chicago, IL 60637, US}
\author{Christopher P. Anderson}
\affiliation{Pritzker School of Molecular Engineering, University of Chicago, Chicago, IL 60637, US}
\author{Alexandre Bourassa}
\affiliation{Pritzker School of Molecular Engineering, University of Chicago, Chicago, IL 60637, US}
\author{David D. Awschalom}
\affiliation{Department of Physics, University of Chicago, Chicago, IL 60637, USA}
\affiliation{Pritzker School of Molecular Engineering, University of Chicago, Chicago, IL 60637, US}
\affiliation{Materials Science Division and Center for Molecular Engineering, Argonne National Laboratory, Lemont, IL 60439, USA}
\author{Giulia Galli}
\email[]{Corresponding author gagalli@uchicago.edu}
\affiliation{Department of Chemistry, University of Chicago, Chicago, IL 60637, USA}
\affiliation{Pritzker School of Molecular Engineering, University of Chicago, Chicago, IL 60637, US}
\affiliation{Materials Science Division and Center for Molecular Engineering, Argonne National Laboratory, Lemont, IL 60439, USA}

\date{\today}% It is always \today, today,

\begin{abstract}

Optically addressable paramagnetic defects in wide-band-gap semiconductors are promising platforms for quantum communications and sensing. The presence of avoided crossings between the electronic levels of these defects can substantially alter their quantum dynamics and be both detrimental and beneficial for quantum information applications. Avoided crossings give rise to clock transitions, which can significantly improve protection from magnetic noise and favorably increase coherence time. However, the reduced coupling between electronic and nuclear spins at an avoided crossing may be detrimental to applications where nuclear spins act as quantum memories.
% Experimentalists also had concerns about the strength of the statement above, so I reformulated it slightly

Here we present a combined theoretical and experimental study of the quantum dynamics of paramagnetic defects interacting with a nuclear spin bath at avoided crossings. We develop a computational approach based on a generalization of the cluster expansion technique,  which can account for processes beyond pure dephasing and describe the dynamics of any solid-state spin-qubits near avoided crossings. Using this approach and experimental validation, we determine the change in nature and source of noise at avoided crossings for divacancies in SiC. We find that we can condition the clock transition of the divacancies in SiC on multiple adjacent nuclear spins states. In our experiments, we demonstrate that one can suppress the effects of fluctuating charge impurities with depletion techniques, leading to an increased coherence time at clock transition, limited purely by magnetic noise. Combined with \textit{ab-initio} predictions of spin Hamiltonian parameters, the proposed theoretical approach paves the way to designing the coherence properties of spin qubits from first principles.
\end{abstract}

\maketitle

%\tableofcontents

\section{Introduction}
In search of solid-state qubits, electron spin defects in wide-band-gap semiconductors have been extensively explored as robust quantum systems offering both long coherence times \cite{Herbschleb2019} and optical read-out \cite{Robledo2011} capabilities for quantum information \cite{Weber} and quantum sensing \cite{Sensing} applications. In order to design optimal qubits, it is critical to understand and control the interaction between the central electronic spin and the nuclear spin bath. The latter determines, at least in part, the coherence time of qubits, as observed in many solid-state color centers \cite{Seo2016, Tyryshkin2012, PhysRevLett.110.027601, PhysRevB.74.035322}, but it also offers a platform for robust multiqubit registers for the development of quantum networks \cite{Abobeih2018, vanderSar2012, PhysRevX.6.021040, Bourassa2020}. 

The description of the interaction between a central spin and the nuclear bath can be particularly challenging when avoided crossings between energy levels of the central spin occur. Avoided crossings give rise to regimes that can be both beneficial and deleterious to the qubit's coherence. For example, operating at the minimum of the spin transition frequency can significantly increase the coherence time when clock transitions arise \cite{Wolfowicz2013, PhysRevLett.105.067602, miao2020universal}.
On the other hand, spin ground state level anticrossings may lead to an undesired increase in longitudinal relaxation rates \cite{PhysRevB.101.155203, PhysRevLett.108.197601} of spin defects. A unified formalism to simulate on equal footing the dynamics of qubits under different conditions is not readily available.

Here we present a theoretical approach to predict the adiabatic dynamics of a qubit interacting with nuclear baths at clock transitions and near ground state level anti-crossings (GSLAC) and validate our predictions experimentally. In particular, we suggest ways to design and optimize the electron-nuclear spin interactions in clock-transition-based quantum devices. Our approach, based on the generalization of the cluster-correlation expansion method \cite{YANG2020168063}, is general and applicable to a broad class of spin defects.

We focus on neutral divacancies (\ch{V_{C}V_{Si}}) in silicon carbide (SiC), which are promising spin qubit candidates \cite{Koehl2011, Klimove1501015, NSon2020, Christle2015}. In particular, the axial divacancy is one of the most commonly studied defect qubits in SiC \cite{Crook2020, Whiteley2019}, with purely axial zero field splitting. The basal divacancy, on the other hand, exhibits both an axial and transverse crystal field splitting component, giving rise to clock transitions at zero magnetic field \cite{Miaoeaay0527, miao2020universal}.
We validate our predictions by carrying out Ramsey and Hahn-echo experiments on the basal divacancy, as well as by comparing our theoretical results to previous measurements \cite{Seo2016}. Using theory and experiment we show that in the presence of strongly coupled nuclear spins, multiple clock transition conditioned on the nuclear spin state may occur. Importantly, we identify the dominant causes of decoherence at avoided crossings and clock transitions, and we discuss the nature of the noise as a function of the magnetic field. Finally, we show experimentally that the impact of the nuclear bath on the qubit dynamics can be isolated by employing a charge depletion technique, which leads to an increase of spin coherence time of clock transition qubits by suppressing electric noise.

The rest of the paper is organized as follows: in Sec. \ref{ThFr} we summarize the proposed theoretical approach. Next, we discuss the decoherence of axial divacancy near GSLAC in Sec. \ref{gslac}. Sec. \ref{clock} contains the Ramsey and Hahn-echo predictions and measurements for basal divacancy near clock transition. We describe a change in nature of the noise near avoided crossings in Sec. \ref{noise}, and in Sec. \ref{protect} we discuss further ways to engineer coherence protections. Finally, we conclude the paper in Sec. \ref{conclusions} discussing the general implications of our findings.

\section{Theoretical Framework} \label{ThFr}
We start by describing a generalization of the cluster-correlation expansion (CCE) method, and we then apply it to specific defects in SiC. 

\begin{figure}
    \centering
    \includegraphics[scale=1]{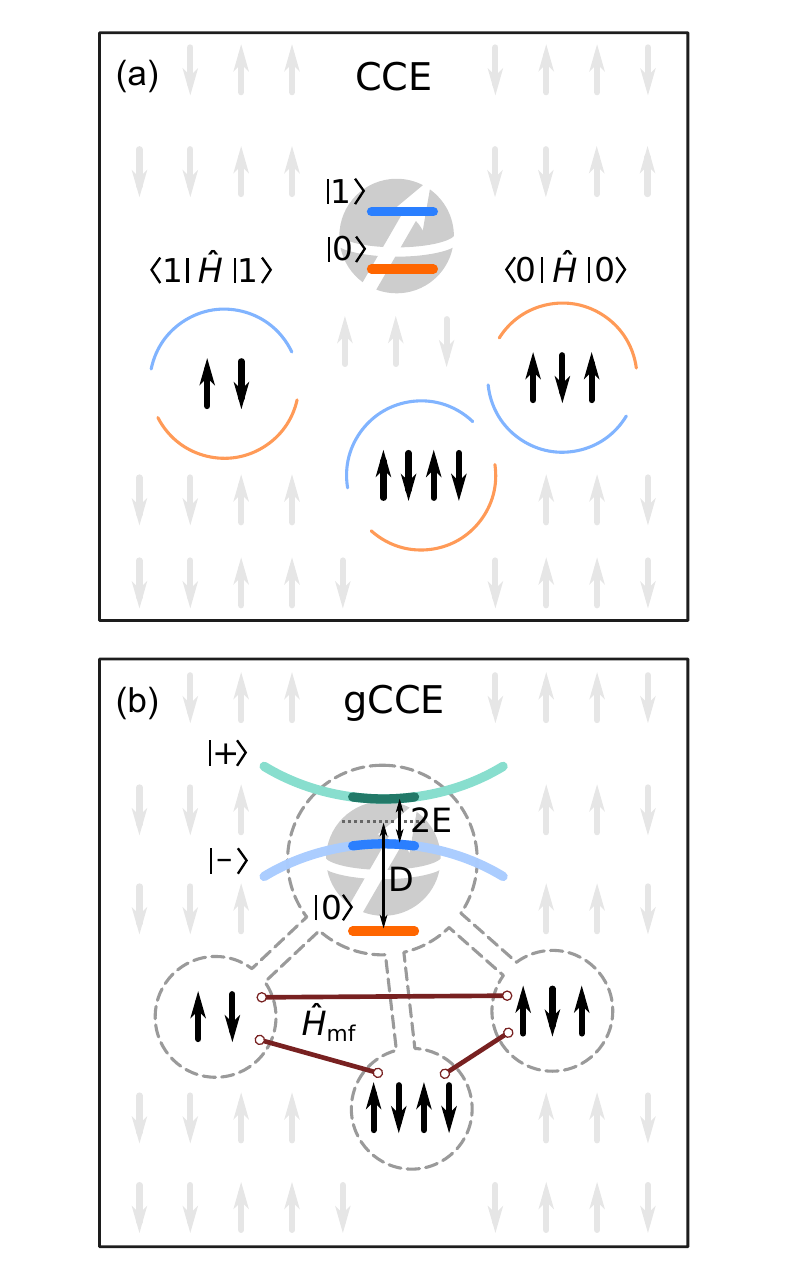}
    \caption{Schematic representation of the conventional (CCE) and generalized (gCCE) Cluster-Correlation Expansion methods. (a) In the conventional CCE the evolution of each bath cluster is conditioned on two qubit levels, $\ket{0}$ and $\ket{1}$, computed using the projected Hamiltonian $\hat H^{(0/1)} = \bra{0/1}\hat H\ket{0/1}$.(b) In the gCCE method each cluster includes the central spin levels (denoted as $\ket{+}$, $\ket{-}$, and $\ket{0}$ for spin-1 with nonzero longitudinal $D$ and transverse $E$ zero field splitting (ZFS)). The interactions between clusters are treated at the mean field level ($\hat H_{\text{mf}}$), using Monte-Carlo sampling of bath states.}
    \label{fig:selling_pic}
\end{figure}

The CCE method is one of the leading approaches for the simulation of quantum dynamics of spin-qubits interacting with a multitude of bath spins \cite{PhysRevB.78.085315, Ye2019, Ma2014, Seo2016}. The method approximates the off-diagonal elements of the qubit density matrix as a product of contributions from independent nuclear spin clusters (Fig.\ref{fig:selling_pic}a). In its conventional formulation, the CCE method is not appropriate to describe clock transitions \cite{PhysRevB.89.045403}, as it does not account for long-range interactions between the bath spins mediated by the central electron spin \cite{2007.00412}; in addition, it cannot describe transitions between spin levels that are close in energy, where the nuclear bath can induce flips of the central spin \cite{Yang_2016}. A generalization of the CCE approach was recently proposed to address the longitudinal relaxation of the \ch{NV^-} center in diamond \cite{YANG2020168063}. In this formulation, each nuclear cluster includes the central spin explicitly, and the cluster contributions are used to compute the change in population of the qubit states. In this work, we both reformulate the generalized CCE method and include a Monte-Carlo bath state sampling (gCCE, Fig.\ref{fig:selling_pic}b), thus enabling the prediction of the dynamics of both the population and the coherence of the central spin at avoided crossings. 

\subsection{Spin defect Hamiltonian}
The Hamiltonian of a central electron spin interacting with a bath of nuclear spins in an external magnetic field can be written as a sum of the central spin Hamiltonian $\hat H_e$, and nuclear Zeeman splitting, hyperfine coupling, and nuclear dipolar coupling terms:

\begin{equation} \label{Total H}
\hat H = \hat H_e - \sum_i \gamma_{n, i} B_z \hat I_{z,i} + \sum_i \textbf{SAI}_i + \sum_{i \neq j} \textbf{I}_i \textbf{P} \textbf{I}_j
\end{equation}

Here $B_z$ is the magnetic field oriented along the $z$ axis, $\gamma_{n, i}$ is the gyromagnetic ratio of the $i$-th nuclear spin, $\textbf{S} = (\hat S_x, \hat S_y, \hat S_z)$ and $\textbf{I}_i = (\hat I_{x, i}, \hat I_{y, i}, \hat I_{z, i})$ denote electron and the $i$-th nuclear spin operators respectively, $\textbf{A}_i$ is the hyperfine coupling tensor of the $i$-th nuclear spin, and $\textbf{P}_{ij}$ is the dipole-dipole coupling between spins $i$ and $j$.

The electron Hamiltonian $\hat H_e$ includes the Zeeman interaction with the external magnetic field and a zero field splitting (ZFS) term with longitudinal ($D$) and transverse ($E$) components:

\begin{equation} \label{H_e}
\hat H_e = - \gamma_e B_z \hat S_z + D \left( \hat S_z^2 - \frac{1}{3}S(S+1)\right) + E(\hat S_x^2 - \hat S_y^2)
\end{equation}

\subsection{Generalization of cluster-correlation expansion method}
In the gCCE method, we compute both diagonal and off-diagonal elements of the density matrix of the central spin $\rho_{ab} = \bra{a} \hat \rho \ket{b}$ as the product of cluster contributions, where $a, b$ denote different spin states (e.g. $m_s = 1, 0, -1$ states for spin 1 systems):

\begin{equation}\label{gCCE_eq}
\rho_{ab} = \tilde \rho_{ab}^{\{0\}}
            \prod_{i} \tilde \rho_{ab}^{\{i\}}
            \prod_{i, j} \tilde 
            \rho_{ab}^{\{ij\}} ...
\end{equation}

Here $\{0\}$ denotes a cluster consisting of a free central spin, $\{i\}$ a cluster including the central spin and nuclear spin $i$, and so on. The maximum size of the cluster in Eq. (\ref{gCCE_eq}) defines the order of the approximation. For example, at first order (gCCE1), only isolated nuclear spins ($\rho^{\{i\}}$) are included in the expansion. At second order (gCCE2), contributions from pairs of nuclear spins ($\rho^{\{i,j\}}$) are added, etc. The cluster contributions are defined recursively:

\begin{equation} \label{cluster_contribution}
    \tilde \rho_{ab}^{\{C\}} = \frac{\bra{a} \hat \rho_{C} (t)\ket{b}} {\prod_{C'} \tilde \rho_{ab}^{\{C' \subset C\}}}
\end{equation}

Where $\rho_{C} (t)$ is the density matrix of the cluster $C$, the superscript $\{C'\}$ indicates all sub-clusters of $C$, including the free central spin sub-cluster $\{0\}$. The density matrix $\rho_{C} (t)$ is computed using the time ordered propagator:
\begin{equation}\label{propagator}
    \hat U_C(t) = \mathcal{T} \left[e^{-i\int_0^t\hat H(\tau) d\tau} \right]
\end{equation}
where the $\mathcal{T}$ is the time ordering operator, and $\hat H(\tau)$ is the time dependent Hamiltonian, which includes only the interactions within a given cluster.

We emphasize that in our formulation we take into account the interaction of spins inside a given cluster with the rest of the bath, thus allowing for different energy splittings within a cluster, and improving the numerical convergence of cluster dynamic simulations \cite{PhysRevB.86.035452}. In particular, we perform calculations with randomly sampled \textit{pure} states of the spin bath, and for each pure state we include the mean-field effect of all the nuclear spins outside a chosen cluster $\hat H_{\text{mf}}$ (Fig.\ref{fig:selling_pic}b). As we show below, the use of the Monte Carlo sampling of bath states is critical to explaining the dynamics of the central spin at avoided crossings. Further details are available in Appendix \ref{app_cluster}.

\section{Decoherence at ground state level anticrossings} \label{gslac}

We start by investigating avoided crossings, and in particular we study the dynamics of the axial \textit{kk}-\ch{V_{C}V_{Si}} defect in 4H-SiC near its GSLAC. The measured ensemble-averaged Hahn-echo coherence times ($T_2$) of this defect reaches 1.3 ms in samples with natural isotopic concentration \cite{Seo2016}. Due to its \ch{C_{3v}} symmetry, the ZFS entering (\ref{H_e}) has only a longitudinal component $D = 1.305$ GHz \cite{Falk2013}, and the qubit levels may be chosen as the $\ket{-1_{z}}$ and $\ket{0_{z}}$ eigenstates of $S_z$, where $z$ is the spin quantization axis.

\begin{figure*}
    \centering
    \makebox[0pt]{
    \includegraphics[scale=1]{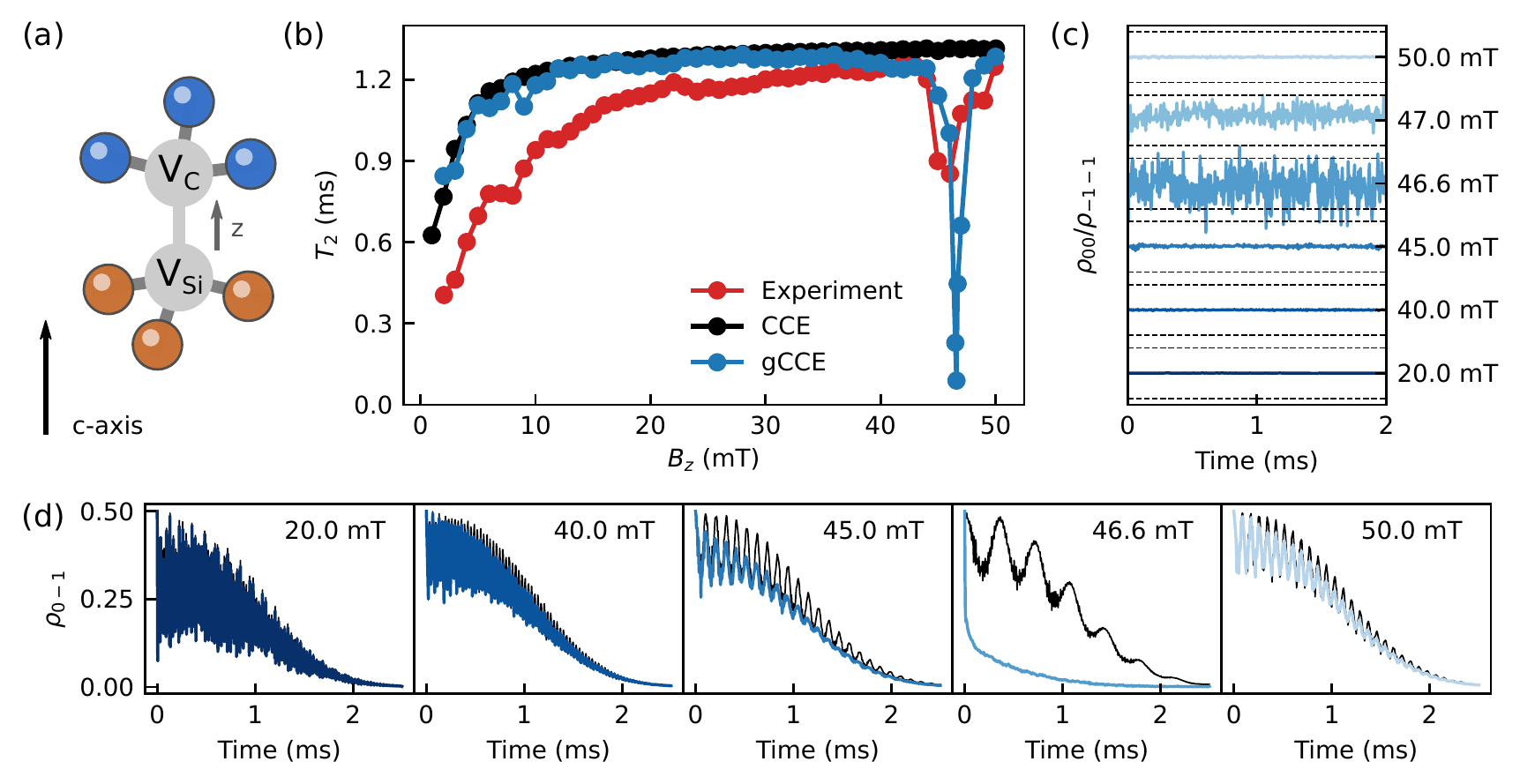}
    }
    \caption{Ensemble Hahn-echo coherence of the axial \textit{kk}-divacancy in 4H-SiC.(a) Schematic representation of the axial \textit{kk}-divacancy. (b) Coherence time $T_2$ (blue dots) as a function of the magnetic field $B_z$. The experimental results (red dots) and conventional CCE predictions (black dots) are from \cite{Seo2016}. (c) The oscillations in the ratio of the diagonal elements of the density matrix of the divacancy as a function of time for various values of the magnetic field. The dashed lines show the $\pm 2\%$ range.(d) The off-diagonal elements of the density matrix for different magnetic fields computed using the gCCE method with Monte-Carlo sampling of the bath states (color) and using the conventional CCE (black).}
    \label{fig:rho_bz_pl2}
\end{figure*}

The coherence time of the \textit{kk}-\ch{V_{C}V_{Si}} divacancy was successfully predicted with the conventional CCE for a wide range of magnetic fields, and the homo-nuclear pair-wise spin flips were found to be the main source of decoherence at strong fields \cite{Seo2016}. However, a significant decrease in the coherence time was observed in experiments when the magnetic field approaches $\sim$ 45 mT; such a decrease is not captured by conventional CCE calculations (Fig.\ref{fig:rho_bz_pl2}b), suggesting a decoherence mechanism beyond pure dephasing.

We performed gCCE calculations with Monte Carlo sampling of bath states (Fig.\ref{fig:rho_bz_pl2}d) and we correctly obtained a local minimum in $T_2(B)$ for $B_z = 46.6$ mT, while reproducing the results of conventional CCE calculations for other values of $B_z$. We note that while previous CCE results \cite{Seo2016} were obtained using a point dipole approximation, here we used accurate hyperfine couplings predicted by \textit{ab initio} calculations (see Methods). This difference in hyperfine couplings accounts for the small discrepancy between CCE and gCCE results observed at small fields.

The origin of the minimum in $T_2(B)$ can be understood by analyzing the populations of different spin levels (Fig.\ref{fig:rho_bz_pl2}c). For most values of $B$, the population of the electron spin levels is constant. However, there are values of $B$ for which the energy difference between the $\ket{\-1}$ and $\ket{0}$ levels is of the same order of magnitude as the hyperfine interaction with nuclear spins. We note that when the energy splitting falls below $\sim$ 100 MHz, significant deviations from pure dephasing starts to occur (see Supplementary Information). In this case, the electron spin experiences large population fluctuations which lead to a significant decrease in coherence time near the GSLAC. Therefore, we conclude that at a GSLAC, the longitudinal relaxation process substantially contributes to decreasing the Hahn-echo coherence time, as observed experimentally.

\section{Decoherence at clock transitions}\label{clock}

As mentioned earlier, the basal divacancy \textit{kh}-\ch{V_{C}V_{Si}} exhibits a clock transition at zero magnetic field. Compared to the Bi:Si donor qubits, studied in Ref.\cite{PhysRevB.89.045403, PhysRevB.92.161403}, the clock transition in this defect arises from ZFS interactions. \textit{kh}-\ch{V_{C}V_{Si}} has $C_{1h}$ symmetry, leading to a nonzero transverse component of the ZFS: $E = 18.4$ MHz \cite{Miaoeaay0527}. Combined with a strong longitudinal splitting ($D = 1.334$ GHz), the ZFS tensor leads to an avoided crossing of electron spin levels at zero magnetic field from which a clock transition emerges.
The frequency of clock transitions is insensitive to magnetic fields to first order, thus increasing protection from the nuclear bath induced decoherence \cite{PhysRevLett.105.067602}, but at the same time it may exhibit an increased sensitivity towards electric noise \cite{PhysRevB.93.024305}.

\begin{figure}
    \centering
    \includegraphics[scale=1]{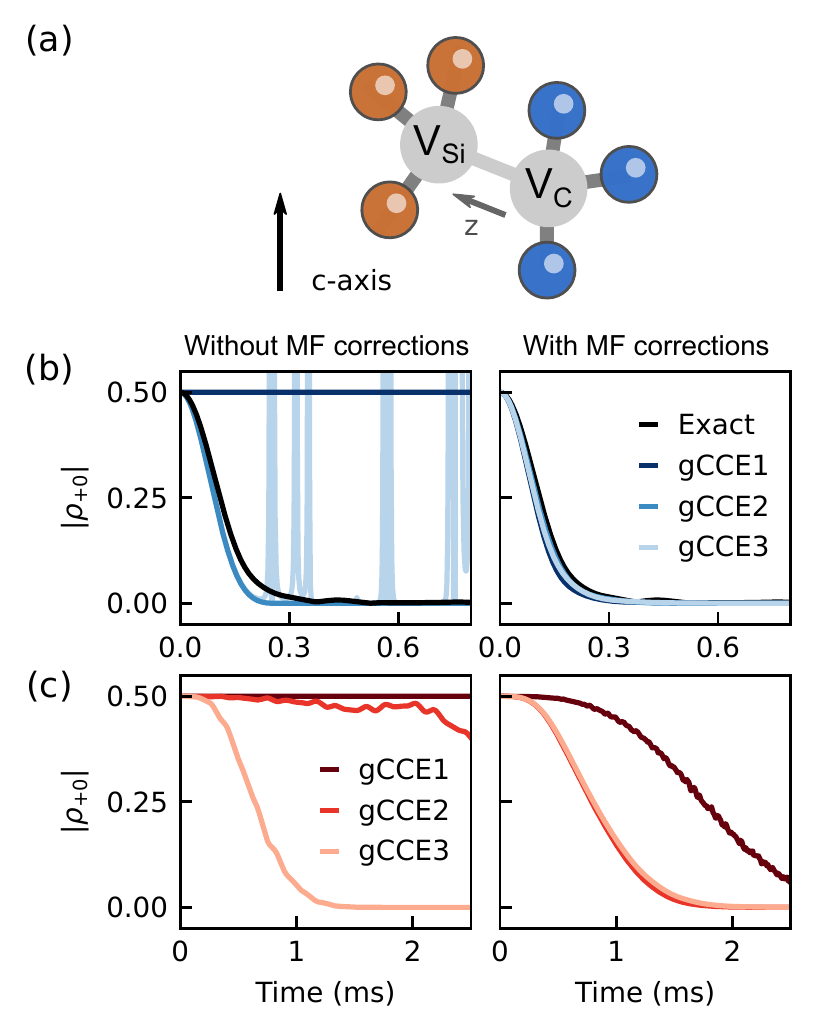}
    \caption{Single defect coherence of the basal \textit{kh}-divacancy in 4H-SiC predicted by different theoretical approximations. (a) Schematic representation of the basal \textit{kh}-divacancy. (b, c) The absolute value of the off-diagonal elements of the density matrix $|\rho_{0+}|=|\bra{0}\hat \rho \ket{+}|$ for one random nuclear configuration corresponding to  Ramsey (b) and Hahn-echo (c) experiments. The results of the gCCE at different orders  (from 1 to 3) without mean field (MF) interactions are shown on the left hand side, with mean field corrections are on the right hand side. The exact solution for a bath of 9 nuclear spins for the Ramsey decay is shown as a black line.}
    \label{fig:pl4order}
\end{figure}

We applied the gCCE method with and without Monte Carlo bath state sampling to reveal the qubit dynamics observed in the Ramsey and Hahn-echo experiments. Figure \ref{fig:pl4order} shows the time evolution of the off-diagonal element of the density matrix of the qubit for one random spatial configuration of nuclear spins. The qubit levels were chosen as the electron Hamiltonian eigenstates $\ket{+} = \frac{1}{\sqrt{2}} (\ket{1_{z}} + \ket{-1_{z}})$ and $\ket{0} = \ket{0_{z}}$ at zero magnetic field, and the qubit was initially prepared in the $\ket{\psi}=\frac{1}{\sqrt{2}}\left(\ket{+} + \ket{0}\right)$ state. For $B \neq 0$, the eigenstate $\ket{+}$ evolves into a different superposition of $\ket{1_{z}}$ and $\ket{-1_{z}}$ (see Supplementary Information). 

We emphasize that considering mean-field effects of nuclear interactions is crucial in order to obtain the correct dynamics of the coherent state. We found that the decay of the observed central spin Ramsey envelope can be accurately described by performing calculations at the gCCE1 level with Monte Carlo bath state sampling (Fig.\ref{fig:pl4order}b). Since gCCE1 simulations do not explicitly include nuclear-nuclear interactions, our results suggest that in the Ramsey experiment the dominant decoherence mechanism is the static Overhauser field generated by nuclear spins, in agreement with Ref. \cite{2007.00412}. On the other hand, we found that for most configurations, gCCE2 is necessary and sufficient to converge the value of the Hahn-echo coherence (Fig.\ref{fig:pl4order}c) time, confirming the significant contribution of nuclear-nuclear interactions.

\subsection{Impact of nuclear spin coupling}

In order to understand how the coupling strength between the central spin and the nuclear spins affects clock transitions, we experimentally investigated three different single \textit{kh}-\ch{V_{C}V_{Si}} divacancy qubits. They are labeled VVA, VVB, and VVC, and they represent configurations with weakly (VVA) and strongly interacting (VVB and VVC) nuclear spins. We obtained theoretical configurations directly comparable with the experimental ones by generating a set of random nuclear spin configurations in the SiC lattice with the same number of strongly interacting nuclear spins as observed in the measured Ramsey fringes. Out of this set we then selected the configurations with a computed value of $T_2^*$ at $B_z = 50$ $\mu$T similar to the measured one (see SI for details).

\begin{figure*}
    \centering
    \makebox[0pt]{
    \includegraphics[scale=1]{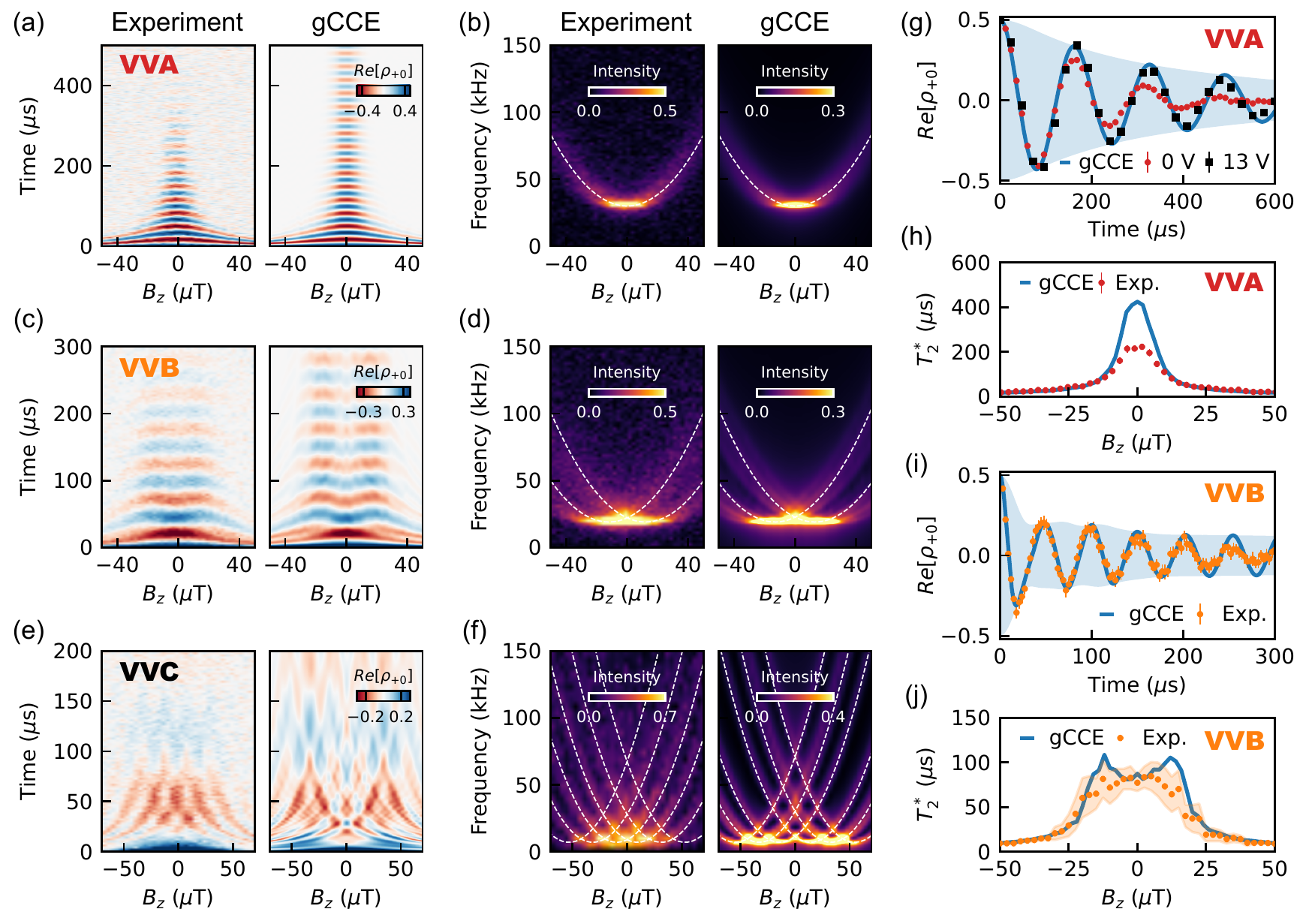}
    }
    \caption{Ramsey interferometry for three experimental \textit{kh}-\ch{V_C V_{Si}} systems.
    (a-f) Ramsey precession and frequency spectrum for a defect with only weakly coupled nuclear spins (VVA, a, b), with one (VVB, c, d), and with three strongly coupled nuclear spins (VVC, e, f). For each defect we show theoretical predictions and experimental results. White dashed lines show the positions of the hyperbolae (see Eq.  \ref{hyperbola}). 
    (g) Measured Ramsey precession of VVA at zero field with (black) or without (red) charge depletion (see text), compared to the theoretical prediction (blue). The shaded area corresponds to the theoretically predicted decay. 
    (h, j) Distribution of $T_2^*$ for VVA (h) and VVB (j) as a function of the magnetic field ($B_z$). Shaded area in (j) shows the error of the fit.
    (i) Measured Ramsey precession of VVB at weak applied magnetic field compared to the theoretical prediction.
    Error bars correspond to 2SD.}
    \label{fig:pl4_fid}
\end{figure*}

We first analyze the VVA configuration, which contains only weakly coupled nuclear spins. Its frequency spectrum, obtained as a Fourier transform of Ramsey fringe oscillations, can be simply represented by one hyperbola (Fig.\ref{fig:pl4_fid}a, b). In the absence of a nuclear bath, the frequency of the clock transition is given by:
\begin{equation}\label{hyperbola}
    \omega - \omega_0 = \sqrt{\gamma_e^2 B_z^2+E^2}
\end{equation}
 
We obtain good agreement between theoretical predictions and the measured Ramsey fringes at small fields, but in the zero field regime the experimentally observed decoherence is significantly faster (Fig.\ref{fig:pl4_fid}h). We found that this apparent discrepancy is due to the electric noise affecting the qubit state, as we explain below.

When operated near a clock transition, the basal divacancy spin becomes first-order insensitive to magnetic fluctuations. However, a first-order sensitivity to electric field fluctuations emerges, due to the linear dependence of the ZFS tensor components on the local electric field. Therefore, the electron spin dephasing time becomes limited by the electric field noise \cite{miao2020universal, PhysRevB.93.024305, Bourassa2020}. In SiC divacancies the electric noise is primarily caused by charge state fluctuations of photoactive impurities, which may undergo charge state transitions under optical excitation \cite{Anderson1225, Wolfowicz2017}, leading to a variation of local electric fields.

In our experiments, we used charge depletion \cite{Miaoeaay0527, Anderson1225} to deactivate photoactive impurities within the optical excitation region, thus substantially reducing the electric field contributions to the ground-state spin dephasing. We applied 13 V across a lithographically patterned capacitor with a 10 $\mu$m gap width. The applied electric field acting on a divacancy located between the capacitor plates ionizes the undesired charge carriers and removes them from photoactive impurities in the proximity of the divacancy. This technique allowed us to isolate the contributions of the magnetic field noise near the clock transition (Fig.\ref{fig:pl4_fid}g) and to perform a meaningful comparison with our theoretical model.

We found that under charge depleted conditions, the measured coherence time is substantially increased and the experimentally observed Ramsey precession at zero field agrees well with the theoretical prediction of the gCCE. Interestingly, in the presence of a weak magnetic field, an agreement between theory and experiment is obtained without applying any charge depletion, indicating that the decoherence rate in this case is not limited by electric noise.

We now turn to lattice configurations with strongly coupled nuclear spins. We first consider the defect labeled VVB, for which we observe a splitting in the frequency spectrum due to the presence of one strongly coupled nuclear spin (Fig.\ref{fig:pl4_fid}c, d). Each of the hyperbolae shown in the figure corresponds to the oscillation frequency of Ramsey fringes of the divacancy, coupled to either the spin-up or spin-down nuclear state. The minimum of each hyperbola occurs when the magnetic field is equal to the hyperfine field of the strongly coupled nuclear spin, $\left| \frac{A_{iz}} {2 \gamma_e} \right| = \left| B_z \right|$ where $A_{iz} = \sqrt{A_{xz}^2+A_{yz}^2+A_{zz}^2}$ \cite{Miaoeaay0527}. By solving this equation, we obtain the hyperfine parameter of the strongly coupled nuclear spin in VVB: $A_{iz} \approx 0.6$ MHz.

In Fig.\ref{fig:pl4_fid}c, d we compare with experiments our theoretical results for a nuclear configuration for which the computed $A_{iz}$ is $0.75$ MHz. We find an excellent agreement for the time evolution and the frequency spectrum. We note that due to the presence of the strongly coupled spin, the Ramsey precession exhibits a fast and a slow decay mode (Fig.\ref{fig:pl4_fid}i), and the full dynamics of the decoherence process may not be described by a single $T_2^*$ (Fig.\ref{fig:pl4_fid}j). However, by initializing the strongly coupled nuclear spin so that it is antiparallel to the external magnetic field, one can eliminate the fast decay mode, and, together with the charge depletion strategy outlined above, one may achieve a substantial increase (by a factor of 5) in the coherence time (see SI).

In the presence of several strongly coupled nuclear spins, further splitting of the frequency spectrum may occur. In this case, the minimum of each hyperbola is located at:
\begin{equation}
    B_z = \sum_\text{strong} {\pm \frac{A_{iz}} {2 \gamma_e}}
\end{equation}
The measured frequency spectrum of the defect labeled VVC contains six separate hyperbolae, suggesting the presence of three strongly coupled nuclear spins with two of them having similar hyperfine parameters (Fig.\ref{fig:pl4_fid}e, f). We identify one nuclear spin with $A_{iz} \approx 1.7$ MHz and two nuclear spins with $A_{iz} \approx 0.6$ MHz.

Hence, we compare theory and experiment using calculations for a nuclear configuration which contains 3 strongly coupled nuclear spins with similar values of the hyperfine constants: $A_{iz} = 1.92,\ 0.65,\ 0.49$ MHz. We obtain a good agreement in both the time and frequency domains. However, due to the complexity of the dynamics, a simple exponential decay cannot reliably characterize the decoherence time of VVC; nevertheless the decoherence occurs on a timescale of 200-300 $\mu$s, both in theory and experiment.

\begin{figure}
    \centering
    \includegraphics[scale=1]{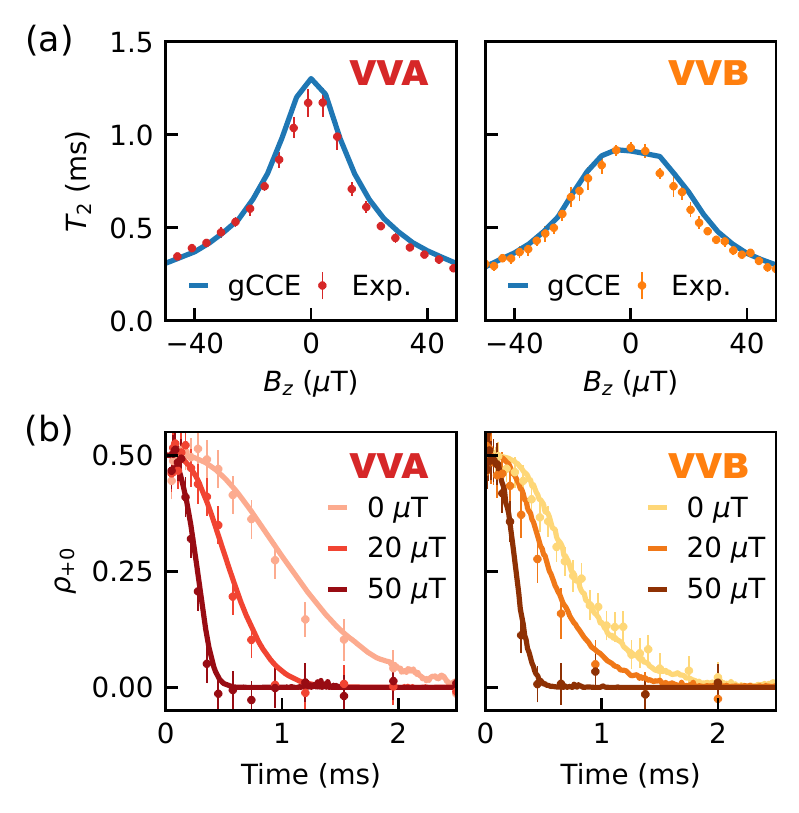}
    \caption{Single defect Hahn-echo coherence time of \textit{kh}-\ch{V_C V_{Si}}. (a) Distribution of $T_2$ for VVA (left) and VVB (right) as a function of the magnetic field $B_z$ (VVA and VVB are the same defects represented in Fig.\ref{fig:pl4_fid}). (b) The Hahn echo decay for three different values of the magnetic field. Solid lines correspond to theoretical predictions, points to experimental measurements. Error bars correspond to 2SD.}
    \label{fig:pl4_he}
\end{figure}

Furthermore, we carried out a study of the Hahn-echo decoherence time for the VVA and VVB defects (Fig.\ref{fig:pl4_he}a, b) and again found excellent agreement between experimental values and theoretical predictions. In VVB, the presence of the strongly coupled nuclear spin leads to a broadening of the coherence time peak compared to VVA, and to a decrease in the maximum of $T_2$ (0.930(14) ms for VVB vs. 1.17(4) ms for VVA). We note that the measured and computed zero-field Hahn-echo coherence times agree even without applying any charge depletion to the sample, suggesting that the electric noise has a minor impact on $T_2$.

\section{Nature of nuclear noise in solid-state qubits}\label{noise}

Having validated the predictions of the gCCE with several experiments, we can now analyze the nature of the nuclear noise in the decoherence processes of the \textit{kh}-\ch{V_{C}V_{Si}} and the axial divacancies. 

\begin{figure}
    \centering
    \includegraphics[scale=1]{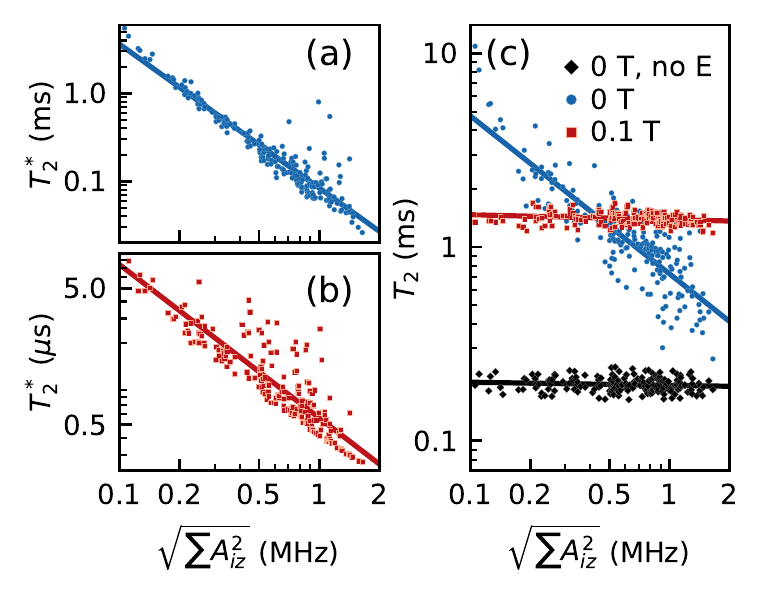}
    \caption{Single defect coherence times $T_2^*$ and $T_2$ of the \textit{kh}-\ch{V_CV_{Si}} at zero and high magnetic field for different bath coupling.
    (a, b) $T^*$ for zero (a) and high (b) field as a function of the square root of the sum of squares of the hyperfine couplings with the bath spins $\sqrt{\sum_{i} A_{iz}^2}$. Significant deviations from the least squares fit (solid lines) are present for systems containing single nuclear spins with high hyperfine coupling, and the coherence decay may not be approximated by a single exponential \cite{PhysRevB.85.115303}. 
    (c) $T_2$ for zero (blue) and high (red) magnetic fields, and for an hypothetical system with $E=0$ MHz at zero field (black). All calculations are performed for SiC with natural isotope concentration for the weakly coupled nuclear bath.}
    \label{fig:pl4suma}
\end{figure}

We computed the coherence time of the \textit{kh}-\ch{V_C V_{Si}} at zero field (0 T), where a clock transition occurs, and at $B_z = 0.1$ T where we expect the basal and axial divacancies to exhibit similar coherence properties. We considered eigenstates of $\hat S_z$ as qubit levels at 0.1 T. We found that at both zero and strong magnetic fields, the Ramsey decay is limited by static thermal noise arising from the entanglement of the qubit with pure states of the bath, which remain unchanged in time \cite{Yang_2016}:
\begin{equation}
 \hat \rho(0) \otimes \sum_B p_B \ket{B}\bra{B} \rightarrow \sum_B p_B \hat \rho_B(t) \otimes \ket{B}\bra{B}   
\end{equation}
Indeed our calculations of $T_2^*$ for the \textit{kh}-\ch{V_CV_{Si}} in the weakly-coupled bath (Fig.\ref{fig:pl4suma}a, b) show that the inhomogeneous coherence time depends on the hyperfine parameters only through the average bath coupling ($\sqrt{\sum_{i} A_{iz}^2}$), as expected in the case of static thermal noise. \cite{PhysRevB.65.205309}.

The nature of noise is different in Hahn-echo experiments, where the $\pi$-pulse removes the static part of the noise dominating the Ramsey decay, and $T_2$ depends only on the dynamical fluctuations of the magnetic field due to nuclear spins flips \cite{Yang_2016}. If the flips are completely random, the decay rates originate primarily from the accumulation of random phases due to dynamical fluctuations; in this case the noise is by definition classical and the variance of the noise distribution is given by $\sum_{i} A_{iz}^2$ \cite{GuClassical}. Therefore we expect the coherence time to vary linearly as a function of $\sqrt{\sum_{i} A_{iz}^2}$ in systems where the noise is classical\cite{PhysRevB.90.064301}. On the other hand, when the back action of the central spin is dominant (i.e. the dynamics of the nuclear bath is strongly influenced by the electron spin state \cite{Yang_2016}), the coherence dynamics deviates from that predicted using classical approximations \cite{Huang2011}. Fig.\ref{fig:pl4suma}c shows $T_2$ of \textit{kk}-\ch{V_C V_{Si}} as a function of $\sqrt{\sum_{i} A_{iz}^2}$ in the zero and strong field regimes. At zero field, $T_2$ varies as $\sqrt{\sum_{i} A_{iz}^2}$, with more than an order of magnitude difference in $\sqrt{\sum_{i} A_{iz}^2}$ between the different configurations. This dependence suggests a stochastic nature of the noise at clock transitions and is consistent with the results reported for bismuth qubits in silicon \cite{PhysRevB.92.161403}. At strong fields, $T_2$ is instead independent on the average coupling to the bath, consistent with the quantum nature of the noise, expected in this regime. These results show that the noise affecting Hahn-echo experiments is different at clock transitions and in the strong field regime, and the transition from classical to quantum noise may be tuned by simply varying the applied magnetic field.

We note that the differences in the nature of the noise is not sufficient to explain why the average value of $T_2$ at the clock transition of the basal divacancy is similar to the one at strong field ($1.15$ ms vs $1.4$ ms). This similarity arises from the combination of two competing effects: strong electron spin back-action, leading in principle to a reduction in coherence time, and the Zeeman splitting of nuclear spins, having instead the opposite effect. We can isolate the effect of the electron spin's strong back-action on coherence time, by considering a hypothetical \textit{kh}-\ch{V_C V_{Si}} with $E = 0$ MHz at zero magnetic field (Fig.\ref{fig:pl4suma}c, 0 T, no E). We found that the $T_2$ of this system is independent from $\sqrt{\sum_{i} A_{iz}^2}$, and the ensemble average $T_2$ is $0.2$ ms, significantly smaller than the one obtained for the clock transition, confirming that the electron back-action is indeed responsible for an increase in the decoherence rate. At strong magnetic field, the Zeeman splitting of the nuclear spins is instead responsible for a decrease in decoherence rates. The splitting can be larger than both the interaction strength between nuclear spins and the hyperfine coupling, leading to the suppression of spin non-conserving flips \cite{PhysRevB.85.115303}. Only the secular pairwise flip-flops of nuclear spins with the same gyromagnetic ratio ($\uparrow\downarrow \leftrightarrow \downarrow\uparrow$) are possible in this regime \cite{PhysRevB.90.241203, Seo2016}, thus greatly reducing the number of possible spin flips and decreasing the decoherence rate. (Fig.\ref{fig:rho_bz_pl2}b).

\section{Engineering qubit protection at a clock transition}\label{protect}
\begin{figure}
    \centering
    \includegraphics[scale=1]{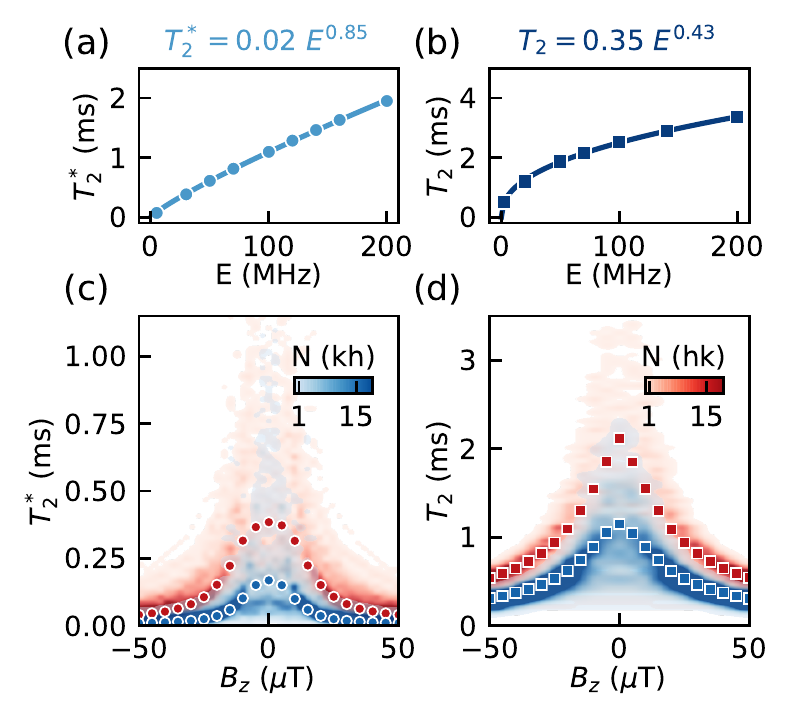}
    \caption{Dependence of the coherence time on the transverse zero-field-splitting (ZFS). (a, b) Coherence times $T_2^*$ (a) and $T_2$ (b) of \textit{kh}-\ch{V_C V_{Si}} with different hypothetical values of transverse ZFS $E$.
    (c, d) The heat map of the single defect $T_2^*$ ($T_2$) for 120 different nuclear spin configurations of both \textit{hk}-\ch{V_C V_{Si}} (red) and \textit{kh}-\ch{V_C V_{Si}} (blue) as a function of magnetic field. The color corresponds to the number of configurations with a given $T_2^*$ ($T_2$) at the given magnetic field. The circles (squares) indicate the ensemble averaged coherence time.}
    \label{fig:E}
\end{figure}

It is interesting to analyze in detail the effect of the magnitude of the transverse component of the ZFS on coherence protection. In order to do so, we investigated how coherence times vary as a function of a hypothetical change in $E$ for the \textit{kh}-\ch{V_{C}V_{Si}}, within a weakly coupled nuclear bath (Fig.\ref{fig:E}(a)).
We find that the ensemble averaged coherence time scales sublinearly as a function of the transverse ZFS ($T_2 \approx 0.34 E^{0.43}$, $T_2^* \approx 0.03 E^{0.61}$; see SI for the distribution of single defect coherence times). Our calculations show that defects with large transverse ZFS will exhibit substantially higher protection from magnetic noise.

% we never talked about hk before, so it is important to show what it is and why we are studying it.
In particular, we predict the coherence time of the \textit{hk}-\ch{V_C V_{Si}} basal divacancy. \textit{hk}-\ch{V_C V_{Si}} has significantly higher transverse ZFS than \textit{kh}-\ch{V_{C}V_{Si}} ($E = 82.0$ MHz), and similar longitudinal ZFS ($D = 1.222$ GHz) \cite{Falk2013}. The total distribution of the $T_2$ and $T_2^*$ as a function of the magnetic field for different spatial configurations of the weakly coupled nuclear bath is shown in Fig.\ref{fig:E}(b, c) for both basal defects. We can see that there is a significant variability in the value of the coherence time at $B=0$. The increase in the transverse ZFS leads both to a significant increase in the maximum value of the coherence time and to an increased robustness towards the external magnetic field. The ensemble average zero field $T_2^* = 380$ $\mu$s of \textit{hk}-\ch{V_C V_{Si}} is predicted to be 2.3 times higher than the one of \textit{kh}-\ch{V_C V_{Si}}, and the $T_2 = 2.12$ ms is found to be increased by a factor of 1.8, in a good agreement with Fig.\ref{fig:E}a. In the presence of a strong field (0.1 T) the ensemble average coherence time for both basal divacancies is the same: $T_2^*=0.7$ $\mu$s, $T_2=1.4$ ms, which confirms that the large transverse ZFS is the main driving force for an increased coherence protection in the \textit{hk}-\ch{V_C V_{Si}}.

Our results for the different basal divacancies show that by engineering high zero field splitting either by selecting different defects, or applying the strain to the system \cite{Falk2014}, one can achieve substantial increase in the coherence time.

\section{Conclusions}\label{conclusions}
Understanding the relation between the electronic structure of spin defects and their coherence properties is pivotal to optimizing the conditions for solid-state qubit applications. 
In this work, we proposed a general computational approach to predict the impact of the dynamics of a central spin on the qubit coherence properties, and we investigated the effect of the nuclear spin bath at avoided crossings for divacancies in SiC. We validated our results with measurements of Ramsey fringes and Hahn-echo coherence times and found excellent agreement between theory and experiments, thus providing a robust strategy to uncover the effect of the interaction of nuclear spins on  solid-state qubits' decoherence over a wide range of applied magnetic fields.

Applying charge depletion \cite{Anderson1225} to electrically improve coherence, we were able to experimentally isolate and elucidate the duality of the nuclear bath impact on clock transitions' dynamics. We discovered that in the presence of strongly coupled nuclear spins, multiple clock transitions in the frequency spectrum of the spin qubit can emerge. We identified and characterized the nuclei with high hyperfine coupling in these systems; the initialization of these nuclear spins should allow one to achieve significantly higher coherence times under applied magnetic fields, while the nuclear-spin dependent spectral features provides guidance for the development of a new class of electron-nuclear two-qubit gates. We found that the effect of weakly coupled nuclear spins can be treated as a stochastic classical noise at the clock transition, and that the total amplitude of the coupling is a good descriptor of the coherence time. We further probed the classical-to-quantum transition of the noise and showed how a tunable back action of the electronic spin emerges with applied magnetic fields.

In sum, the theoretical approach developed in this work allows for the predictions of the dynamical and decoherence properties of solid-state qubit systems with complex spin degrees of freedom, in the presence of clock transitions and over a wide range of magnetic fields. Here we focused on specific defects in SiC, but the approach is general and can be used to describe several other systems of interest such as molecular qubits with multiple clock transitions \cite{Shiddiq2016}, bismuth donor spin qubits in silicon \cite{Wolfowicz2013}, and other  solid-state centers with complex energy structure. Combined with \textit{ab initio} predictions of the spin Hamiltonian parameters \cite{Ghosh2019}, our approach paves the way to optimizing and eventually designing the coherence properties of spin qubits yet to be explored experimentally. 

\section{Acknowledgements}
This work made use of resources provided by the University of Chicago’s Research Computing Center, the UChicago MRSEC (NSF DMR-1420709) and Pritzker Nanofabrication Facility, which receives support from the SHyNE, a node of the NSF’s National Nanotechnology Coordinated Infrastructure (NSF ECCS-1542205).
MM.O.and G.G. were supported by AFOSR FA9550-19-1-0358, H.M. by the UChicago MRSEC (NSF DMR-1420709). K.C.M., J.P.B., C.P.A., A.B., and D.D.A. were supported by AFOSR FA9550-19-1-0358, DARPA D18AC00015KK1932, and ONR N00014-17-1-3026.

M.O. developed the model and conducted the dynamics calculations. H.M. performed the ab initio DFT calculations.
M.O., K.C.M., and J.P.B. designed the experiments. K.C.M. and J.P.B. performed the experiments.
M.O., K.C.M., and J.P.B. analyzed the data.
C.P.A. fabricated the sample.
K.C.M. and J.P.B. developed the confocal microscope setup, with the assistance of A.B.
D.D.A. advised on all experimental efforts.
G.G. advised on theoretical efforts and supervised the project. 
All authors contributed to the writing of the manuscript.

\appendix

\section{Experimental Measurements of the \textit{kh}-\ch{V_C V_{Si}} coherence properties}
Our 4H-SiC sample consists of a 20 $\mu$m high-purity i-type SiC layer epitaxially grown on a 4$^{\circ}$ off-axis miscut of the Si face of a high-purity semi-insulating SiC substrate (serial number A3177-14, Norstel AB). Neutral divacancies are uniformly produced throughout the epitaxial i-type 4H-SiC by electron irradiation with 2-MeV electrons at a dose of $3 \times 10^{12}$ e/cm$^2$ followed by annealing at 850 $^{\circ}$C for 30 min in Ar. A coplanar capacitor structure with a 10 $\mu$m gap width and a wire with 10 $\mu$m width made of Ti/Au are then patterned on the sample surface using electron beam lithography. Samples are cooled to 5 K in a closed-cycle cryostat (Cryostation s100, Montana Instruments).

The confocal microscope consists of a 905 nm excitation laser (QFLD-905-200S, QPhotonics) for off-resonant spin initialization, as well as a narrow-line tunable laser (DL pro, TOPTICA Photonics) for resonant spin readout. We focus these excitation beams through a microscope objective (LCPLN100XIR, Olympus). We detect the filtered optical signal with >80\% quantum efficiency using a low-jitter, low-dark count superconducting nanowire single-photon detector (SNSPD; Opus One, Quantum Opus). Electrical pulses from the SNSPD are counted using a data acquisition module (PCI-6259, National Instruments).

We drive the spin transition $\ket{0} \leftrightarrow \ket{+}$ using signal generators (SG396, Stanford Research Systems) modulated by an arbitrary waveform generator (HDAWG8, Zurich Instruments). The output of the signal generator is routed to the on-chip wire, which produces ac magnetic fields. Vector control of the magnetic field is obtained using a three-axis electromagnet outside the cryostat. 

% We reject laser scatter and background emission while retaining \textit{kh} divacancy phonon-sideband emission in the range of 1090 nm to 1300 nm by filtering the emission from the sample with a tunable longpass filter (TLP01-1116-25x36, Semrock), a fixed longpass filter (LP02-1064RE-25, Semrock), a fixed shortpass filter (89-676, Edmund Optics), and a notch filter (NF1064-44, Thorlabs)

\section{Calculation of cluster contributions in the gCCE}\label{app_cluster}
In order to evaluate the density matrix $\rho_{C} (t)$ in equation (\ref{cluster_contribution}) we compute the evolution of the initial density matrix of a given cluster as:

\begin{equation} \label{cluster_dm}
\hat \rho_{C} (t) = \hat U_C \hat \rho_{C} (0) \hat U_C^\dagger
\end{equation}

Using time ordered propagator $\hat U_C$ \ref{propagator}.

The Hamiltonian used to model Ramsey experiments does not depend on time $\hat H(\tau) = \hat H_C$, and the propagator is trivial:

\begin{equation} \label{fid_propagator}
\hat U_C(t) = e^{-i\hat H_C t}
\end{equation}

The Hamiltonian $\hat H_C$ is equal to the system Hamiltonian (\ref{Total H}), which contains only the central spin and a given cluster of nuclear spins:

\begin{equation}\label{H_C}
\hat H_C = \hat H_e +
 \sum_{i\subset C} \textbf{SAI}_i -\sum_{i\subset C} \gamma_n B_z \hat I_{z,i} + \sum_{i\neq j\subset C} \textbf{I}_i \textbf{P} \textbf{I}_j
\end{equation}

Under the dynamical decoupling to the qubit, by assuming ideal instantaneous control pulses, we can write the propagator as follows:

\begin{equation} \label{dd_propagator}
\hat U_C(t) = 
\mathcal{T} \left[e^{-i \hat H_C \tau} e^{-i \sigma_{\{x, y, z\}} \frac{\phi}{2}} e^{-i \hat H_C \tau}\right]^N
\end{equation}

where $\sigma_{\{x,y,z\}}$ is one of the Pauli matrices (depending on the type of the pulse), spanned by two qubit levels, $\tau$ is the delay between pulses, $\phi$ is the angle of rotation (equal to $\pi$ for CPMG, XY4 sequences \cite{deLange60}; it may be varied to represent more complicated schemes \cite{PhysRevLett.124.220501}) and $N$ is number of pulses. For example, the propagator used to model Hahn-echo experiments with a $\pi$ rotation about the $x$ axis is defined as follows:

\begin{equation} \label{he_propagator}
\hat U_C^{HE}(t) = e^{-i\hat H_C \tau}e^{-i\sigma_{x} \frac{\pi}{2}} e^{-i \hat H_C \tau}
\end{equation}

The Pauli matricies for qubit levels $\ket{0}$ and $\ket{1}$ are defined as:

\begin{align} \label{sigma}
\sigma_{x} = (&\ket{0}\bra{1} + \ket{1}\bra{0}) \\
\sigma_{y} = i(&\ket{1}\bra{0} - \ket{0}\bra{1}) \\
\sigma_{z} = (&\ket{0}\bra{0} - \ket{1}\bra{1})
\end{align}

When using Monte-Carlo sampling of the bath states, we perform the CCE calculations for each pure bath state separately. In the pure bath state, each nucleus is initialized in the spin up or spin down state, and in the mixed state each nuclear spin has a classical probability of being in one of the two states.
We define the density matrix elements of the central spin as follows:

\begin{equation}\label{mixed_bath}
\hat \rho_{ab}(t) = \sum_B p_J \hat \rho_{ab}^J(t)
\end{equation}

where the elements of the density matrix $\hat{\rho}_{ab}$ are written as a summation over pure bath states $J$, with elements $\hat{\rho}^J_{ab}$ and probability $p_J$. In the case of a completely randomized bath (the density matrix of each nuclear spin is equal to $I/2$), the probability $p_J$ is the same for all pure bath states. At the typical temperatures of the experiment ($\ge 4$ K) the nuclear bath can be considered completely randomized.

The procedure used to evaluate the density matrix elements is the following. First, we generate a set of random pure bath states. For each bath state we perform CCE calculations to obtain the electron spin density matrix. Finally, we compute the density matrix elements for the mixed bath state from equation (\ref{mixed_bath}), and verify the convergence of density matrix elements $\rho_{ab}(t)$ with respect to the number of generated bath states (see Supplementary Information). 

We add the mean field effect of the bath spins outside a given cluster, by adding the $\hat H_\text{mf}$ term into the cluster Hamiltonian (\ref{H_C}). The mean field term is defined as:

\begin{equation} \label{H_mf}
\hat H_\text{mf} = \sum_{i\not\subset C}\left[ A_{zz} \langle I_{z,i} \rangle \hat S_z + \sum_{j \subset C}{ P_{zz} \langle I_{z,i} \rangle \hat I_{z,j}} \right]
\end{equation}

where $\langle I_{z,i} \rangle = \pm \sfrac{1}{2}$, is the projection of the nuclear spin in the $z$ direction. The sign depends on the initial state of the nuclei $i$ in the given random bath state.

\section{Calculations of hyperfine coupling of nuclear spins}
The inhomogeneous coherence time $T_2^*$ is directly related to the hyperfine couplings. Under weak magnetic fields, the nuclear spin flips can be induced by both the hyperfine coupling and dipolar-dipolar interactions between nuclear spins \cite{PhysRevB.85.115303}; hence accurate predictions of the hyperfine parameters are necessary to correctly compute the Hahn-echo coherence time as well.

We performed \textit{ab initio} Density Functional Theory (DFT) calculations to predict hyperfine coupling constants for nuclear spins at distances up to 1 nm from the defect and we used the point dipole approximation for spins at larger distances. DFT calculations using the PBE functional were carried out with the GIPAW code \cite{GIPAW} using single-particle wavefunctions obtained with the Quantum Espresso code \cite{Giannozzi2009}. Wavefunctions are represented on a plane-wave basis with a kinetic energy cutoff of 40 Ry. GIPAW pseudopotentials \cite{Ceresoli} are used to model electron-ion interactions. Divancancies are modeled with $9 \times 5 \times 2$ orthorhombic supercells containing 1438 atoms and the Brillouin zone is sampled with the $\Gamma$ point only.

We define a weakly coupled bath as a bath in which the nuclear spins do not change the energy splitting of the defect. We impose a cutoff of the hyperfine couplings of $A_{zz} < 1$ MHz present in the weakly coupled bath, which is of typical order of magnitude compared to strongly coupled nuclear spins in NV center \cite{Yun_2019, PhysRevB.85.134107}. The ensemble dynamics throughout the text is shown for the weakly coupled bath.

\providecommand{\noopsort}[1]{}\providecommand{\singleletter}[1]{#1}%

\newpage

\newpage

\newpage

\newpage

\newpage

\newpage

\newpage

\end{document}